\documentclass[sigconf,manuscript]{acmart}
\usepackage{subfigure}
\AtBeginDocument{%
  \providecommand\BibTeX{{%
    \normalfont B\kern-0.5em{\scshape i\kern-0.25em b}\kern-0.8em\TeX}}}

\setcopyright{iw3c2w3g}
\copyrightyear{2021}
\acmYear{2021}
\acmDOI{}

\acmConference[KidRec'21]{KidRec '21: 5th International and Interdisciplinary Perspectives on Children \& Recommender and Information Retrieval Systems (KidRec) Search and Recommendation Technology through the Lens of a Teacher- Co-located with ACM IDC 2021}{June 26, 2021}{Online Event}
\acmBooktitle{KidRec '21: 5th International and Interdisciplinary Perspectives on Children \& Recommender and Information Retrieval Systems (KidRec) Search and Recommendation Technology through the Lens of a Teacher- Co-located with ACM IDC 2021,
  June 26, 2021, Online Event}
\acmPrice{}
\acmISBN{}

\begin{document}

\title{CASTing a Net: Supporting Teachers with Search Technology}


\author{Garrett Allen}
\email{GarrettAllen@u.boisestate.edu }
\affiliation{%
  \institution{Department of Computer Science-- PIReT -- Boise State University}
  \country{USA}
  \postcode{83725}
  }

\author{Katherine Landau Wright}
\email{katherinewright@boisestate.edu}
\affiliation{%
  \institution{Department of Literacy, Language and Culture -- Boise State University}
  \country{USA}
  \postcode{83725}
}

\author{Jerry Alan Fails}
\email{jerryfails@boisestate.edu}
\affiliation{%
  \institution{Department of Computer Science -- Boise State University}
  \country{USA}
  \postcode{83725}
  }
  
\author{Casey Kennington}
\email{caseykennington@boisestate.edu}
\affiliation{%
  \institution{Department of Computer Science -- Boise State University}
  \country{USA}
  \postcode{83725}
  }
\author{Maria Soledad Pera}
\orcid{0000-0002-2008-9204}
\email{solepera@boisestate.edu}
\affiliation{%
  \institution{Department of Computer Science-- PIReT -- Boise State University}
  \country{USA}
  \postcode{83725}
  }

\renewcommand{\shortauthors}{Allen, et al.}

\begin{abstract}
Past and current research has typically focused on ensuring that search technology for the classroom serves children. In this paper, we argue for the need to broaden the research focus to include teachers and how search technology can aid them. In particular, we share how furnishing a behind-the-scenes portal for teachers can empower them by providing a window into the spelling, writing, and concept connection skills of their students.
\end{abstract}



\maketitle

\section{Search Technology to Support Teachers in the Classroom}

Technology, in particular search technology, is now part of many traditional classrooms. The use of search technology has become even more prevalent in non-traditional classroom settings, as instruction has largely gone remote in response to the ongoing COVID-19 pandemic \cite{kim2020learning}. Research thus far has looked into how children engage and use search tools, the limitations they face, and how to address them via algorithmic functionality that complements mainstream search tools or through search tools tailored specifically to children \cite{danovitch2019growing,gossen2016search,landoni2019sonny,milton2020korsce}. However we have observed that in the classroom context, one of the main concerned parties in this process has received less attention: the teacher. 

Teachers are an integral part of a child's learning journey, serving as a guide and instructor. As the presence of search technology in classrooms grows, the role of the teacher must expand to encompass it. Teachers are in charge of seamlessly incorporating search technology into the classroom environment, but just because they use search technology for everyday online inquiries outside the classroom, it does not mean that they have the specific know-how to complete this integration \cite{ekstrand2020enhancing}. More importantly, students' interactions with search tools -- beyond just the outcomes from online inquiry tasks -- can offer teachers another set of indicators of student's learning. Such interactions can give valuable information to assist teachers in their understanding of the challenges that children encounter while learning, which in turn allows teachers to better scaffold classroom instruction \cite{azpiazu2016finding,holstein2017intelligent}. 

\citet{azpiazu2017online} explore existing search tools that allow teachers to either seek for learning materials to use or to monitor their students' progress. In the case of the latter, the focus is on monitoring the reading level of students, but not other areas of development such as spelling, e.g., Biblionasium\footnote{\url{https://www.biblionasium.com/\#tab/content-spring-picks}}. By focusing only on reading levels, there is a large missed opportunity for teachers to gain a better idea of their students' capabilities, and how to tailor instruction to help advance those capabilities. In this paper, we share our vision for a teacher portal that we argue can supply the technical functionality to enable teachers to assess students' development and offer the necessary support for the varied student capabilities present in their classrooms, physical or remote. Informed by the outcomes of our ongoing research project focused on children's search in the classroom, we discuss the help currently offered by the \textbf{teacher portal} that is part of the Child Adaptive Search Tool (\textbf{CAST})\footnote{\url{https://cast.boisestate.edu/about/}} and how it could be expanded. Along the way, we discuss potential challenges around making this portal a reality, and future research directions related to this topic involving the natural language processing, information retrieval, human-computer interaction, and education and literacy communities.

\section{A Teacher Portal for CAST}
\label{sec:portal}

Teachers have expressed an openness to monitoring tools that allow them to efficiently allocate time among their students \cite{holstein2017intelligent}. They also turn to online tools that can help them with classroom instruction \cite{azpiazu2017online}. A prototype for such a tool is described by \cite{azpiazu2016finding}. This prototype, referred to as a teacher portal, includes tracking for children's reading and comprehension advancements, empowering teachers to focus on struggling students. Motivated by the discussions in \cite{holstein2017intelligent,azpiazu2016finding}, we augment CAST functionality with a teacher portal to give teachers access to settings that will alter the search configuration, enabling teachers to better tune CAST to their students. Further, the portal displays a simple overview of the search data generated by a given classroom, which is of interest to teachers. 

\begin{figure}[t]
    \centering
    \subfigure[Settings toggles\label{subfig:CAST-tp-toggles}]{
        \includegraphics[height=0.3\textheight]{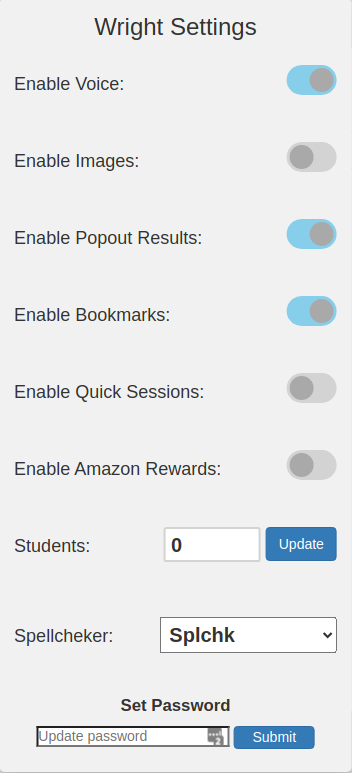}
    }\hspace{10mm}
    \subfigure[Query view\label{subfig:CAST-tp-query}]{
        \includegraphics[width=0.725\textwidth]{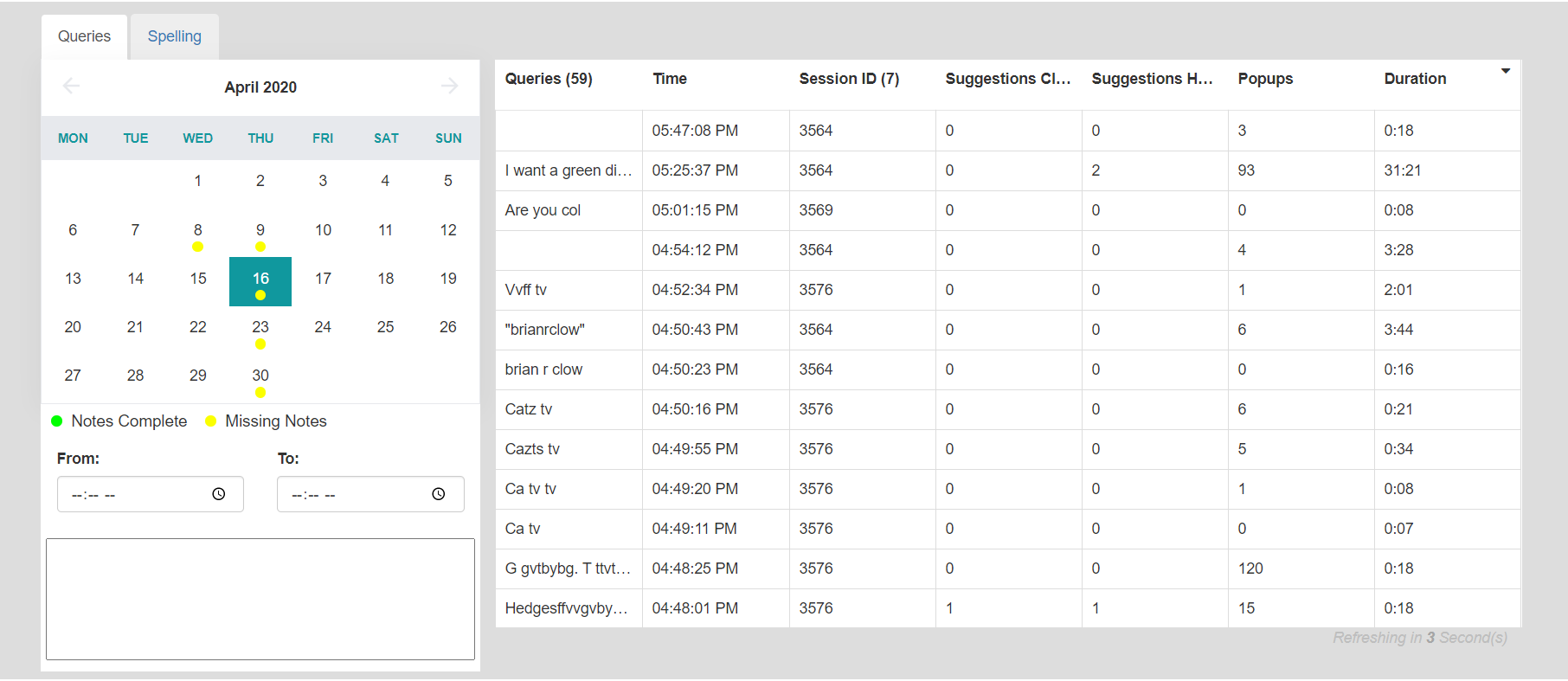}
    }
    \caption{CAST Teacher Portal}
    \label{fig:CAST-tp}
\end{figure}

Currently, the CAST teacher portal contributes two major pieces of functionality: (1) a way to customize the CAST search interface (see Figure \ref{subfig:CAST-tp-toggles}) and (2) a centralized location to view and analyze key pieces of data collected during children's search activities (see Figure \ref{subfig:CAST-tp-query}). Teachers can toggle a number of settings from within the portal to customize the search interface, ensuring CAST is configured for the abilities of the class. For example, CAST contains a spellchecker, KidSpell \cite{downs2020kidspell}, that will circle misspelled words and display a drop down of suggested corrections, which will show an image next to each suggestion and be read aloud by a virtual voice assistant. The images and voice assistant can be toggled on and off from within the portal. The image toggle will also control whether images display next to search results on the result page. Additional options include enabling individual search sessions, allowing searchers to bookmark individual results, and toggling whether hovering over a result will produce a small pop-up window with a preview of the page behind the link.

While children interact with CAST, data related to what the children search for, what queries they use, and how they formulate those queries is recorded\footnote{No identifying information is collected or stored in accordance with the Children's Online Privacy Protection Act and the General Data Protection Regulation.}. If any spelling errors occur during the query, the misspelled word along with the suggestions displayed are recorded. Any correction options that are highlighted (hovered over) or clicked are also recorded. Through a video replay system, teachers can go back and watch each query performed. The teacher portal also includes a simple way to access past queries via a monthly calendar interface with customizable time range searching. This allows teachers to see an overview -- even in real time -- of all queries made by the students in their class. As a part of the calendar, teachers can attach extra notes to each days queries with thoughts or outcomes related to what occurred during that days activities.

\section{A Wide Net of Opportunities}
\label{sec:opps}

In its current form, the CAST teacher portal allots access to data that affords teachers with many opportunities to reinforce their students education. New information is better retained when students can connect new content to existing knowledge \cite{driscoll2005psychology}. Understanding how a child is integrating new knowledge with previously-learned content can help teachers identify misconceptions and areas for reteaching. Examining the queries children use, presented in a list format (see Figure \ref{subfig:CAST-tp-query}), along with the replay system within the CAST teacher portal might give teachers a glimpse into this otherwise invisible cognitive process. For instance, if a child is researching hedgehogs, a teacher would hope they could connect this with existing knowledge about animal survival and thus compose queries such as ``hedgehog habitats'' and ``hedgehog food''. However, if the child searches for ``hedgehog houses'' or even ``Sonic the hedgehog'', the teacher would know that further instruction is needed. 

The replay system can also present insights on children's spelling. One way to evaluate a child's spelling skills is to see how they spell words in their writing \cite{helman2019words}. However, when children know someone, like their teacher, will be reading their work, they are more likely to look up words or default to words they are confident they can spell. This has limited use to teachers, as it is in students' spelling errors that they can identify the patterns they still need to learn. The portal could give teachers a unique view into how students spell in more natural settings, which can have implications for spelling instruction. 

Children will be able to spend less cognitive load on how to spell a word as their skills improve, allowing more attention to be put on forming queries. It has long been established that children need direct instruction and modeling for writing in specific genres \cite{lattimer2003thinking} -- writing search queries should be no exception. There are genre conventions that are necessary for a child to learn in order to be an effective and efficient user of online information. While some students will learn these conventions at home through exposure (such as watching their parents conduct searches) most need explicit instruction. As a means of search as learning \cite{collins2017search}, if a teacher can see their class's search queries and know which children are searching for ``hedgehog habitats'' versus asking ``where do hedgehogs live?'', that teacher will be able to proffer targeted intervention for the children who need it most.

\section{Mending the Net by CASTing an Eye to the Future}

None of the portals explored by \citet{azpiazu2017online} tailor retrieved documents to the current level of education of a user outside their reading skills. Expanding the CAST teacher portal to include the age and background knowledge of students will present teachers with multiple points of interest to review when considering what assistance students may need. For instance, if a particular student consistently selects resources from a ranked list that is above their reading level, leading to the student struggling with gaining the intended knowledge, a teacher can see these trends via the portal and issue the guidance for the student to select results that are easier to read.

In its current form, CAST does not offer query suggestions or query reformulation assistance. The addition of these features, with similar interaction recording to the spellchecking described in \ref{sec:portal}, will benefit teachers by allowing them a glimpse into whether their students are developing the necessary skills to write effective queries. A query suggestion can be seen as a model, as mentioned in Section \ref{sec:opps}, for an effective query. If a student is consistently clicking on suggestions, but the form of their original queries follows consistent patterns, this can be seen as an indicator for the teacher that the student is not developing the appropriate writing skills.

Many of the examples thus far have been oriented around a single student scenario, but a class is made of many students. Supporting students as individuals is valuable, as is the ability to assess the distribution of skills throughout the entirety of a class. To this end, extending the CAST teacher portal to include aggregation analysis, could give teachers a big picture overview. Connecting the measures provided by such an aggregation to the relevant educational standards will produce a further layer of assistance for teachers.

While the CAST teacher portal offers many unique possibilities for teachers to champion their students' development, the portal still faces a substantial challenge to overcome: adoption. A system can be the perfect solution to a particular problem, but if no one knows about it, it will not get used. In order for the previously mentioned benefits to come to fruition, the CAST teacher portal has to make its way into the hands of teachers. Another limitation we see is that related to data sharing. CAST is part of an ongoing research project, yet any data gathered as a result of design, evaluation, and deployment cannot be shared with the research community at large due to children falling under online privacy rules such as the Children's Online Privacy Protection Act and the General Data Protection Regulation. Therefore, while teachers can benefit greatly from CAST data generated by each of their students, the collection of such personally identifying information is prohibited by the aforementioned regulations. Navigating these seemingly contradicting requirements serves as one avenue of future investigation.

The work presented herein represents an initial foray into better supporting teachers as they guide their students in learning. To further ensure teachers are empowered in their classrooms through search technology, their direct input is invaluable; we envision conducting user studies involving teachers with the goal of identifying avenues of research that have not been considered to date.

\begin{acks}
Work funded by NSF Award \# 1763649.
\end{acks}

\bibliographystyle{ACM-Reference-Format}
\bibliography{References}

\end{document}